\documentclass[aps,prl,reprint,showpacs,floatfix,footinbib,superscriptaddress,longbibliography]{revtex4-1}

\usepackage{graphicx}
\usepackage{color}
\usepackage{amsmath,amssymb}
\usepackage{natbib}

\allowdisplaybreaks
\begin{document}

\title{Giant Spin Lifetime Anisotropy in Graphene Induced by Proximity Effects}

\author{Aron W. Cummings}
\email{aron.cummings@icn2.cat}
\affiliation
{
	Catalan Institute of Nanoscience and Nanotechnology (ICN2), CSIC and BIST,
	Campus UAB, Bellaterra, 08193 Barcelona, Spain
}

\author{Jose H. Garcia}
\affiliation
{
	Catalan Institute of Nanoscience and Nanotechnology (ICN2), CSIC and BIST,
	Campus UAB, Bellaterra, 08193 Barcelona, Spain
}

\author{Jaroslav Fabian}
\affiliation
{
	Insitute for Theoretical Physics,
	University of Regensburg,
	93040 Regensburg, Germany
}

\author{Stephan Roche}
\email{stephan.roche@icn2.cat}
\affiliation
{
	Catalan Institute of Nanoscience and Nanotechnology (ICN2), CSIC and BIST,
	Campus UAB, Bellaterra, 08193 Barcelona, Spain
}
\affiliation
{
	ICREA, Instituci\'{o} Catalana de Recerca i Estudis Avan\c{c}ats,
	08070 Barcelona, Spain
}
\date{\today}

\begin{abstract}
We report on fundamental aspects of spin dynamics in heterostructures of graphene and transition metal dichalcogenides (TMDCs). By using realistic models derived from first principles we compute the spin lifetime anisotropy, defined as the ratio of lifetimes for spins pointing out of the graphene plane to those pointing in the plane. We find that the anisotropy can reach values of tens to hundreds, which is unprecedented for typical 2D systems with spin-orbit coupling and indicates a qualitatively new regime of spin relaxation. This behavior is mediated by spin-valley locking, which is strongly imprinted onto graphene by TMDCs. Our results indicate that this giant spin lifetime anisotropy can serve as an experimental signature of materials with strong spin-valley locking, including graphene/TMDC heterostructures and TMDCs themselves. Additionally, materials with giant spin lifetime anisotropy can provide an exciting platform for manipulating the valley and spin degrees of freedom, and for designing novel spintronic devices.
\end{abstract}

\pacs{72.80.Vp, 72.25.Rb, 71.70.Ej}

\maketitle

\textit{Introduction}. Following the discovery of graphene in 2004 \cite{Novoselov2004}, a host of other two-dimensional (2D) materials have been synthesized and studied, each demonstrating unique properties and showing promise for technological applications \cite{Robinson2015}. Currently, there is a great deal of interest in layered heterostructures of these materials \cite{Geim2013, Novoselov2016}, where the combined system might be engineered for specific applications \cite{Ferrari2015} or might enable the exploration of new phenomena \cite{Li2016, Costanzo2016}. In the field of spintronics, graphene has exceptional charge transport properties but weak spin-orbit coupling (SOC) on the order of 10 $\mu$eV \cite{Gmitra2009}, which makes it ideal for long-distance spin transport \cite{Banszerus2016, Drogeler2016, Roche2015} but ineffective for generating or manipulating spin currents. To advance towards spin manipulation, recent work has focused on heterostructures of graphene and magnetic insulators \cite{Yang2013, Wang2015a, Dushenko2016, Hallal2017, Leutenantsmeyer2017} or strong SOC materials such as transition metal dichalcogenides (TMDCs) and topological insulators \cite{Jin2013, Kaloni2014, Gmitra2015}. The SOC induced in graphene by a TMDC could enable phenomena such as topological edge states \cite{Gmitra2016} or the spin Hall effect \cite{Avsar2014, Garcia2017, Torres2017}.

To this end, a variety of recent experiments have explored spin transport in graphene/TMDC heterostructures \cite{Avsar2014, Wang2015b, Yang2016, Wang2016, Omar2017, Yan2016, Dankert2017}. Magnetotransport measurements revealed that graphene in contact with WS$_2$ exhibits a large weak antilocalization (WAL) peak, revealing a strong SOC induced by proximity effects \cite{Wang2015b, Yang2016, Wang2016,Yang2017}. Fits to the magnetoconductance yielded spin lifetimes $\tau_s \approx 5$ ps, which is two to three orders of magnitude lower than graphene on traditional substrates \cite{Drogeler2016, Kamalakar2015}. It was later asserted that after the removal of a temperature-independent background, $\tau_s$ becomes at most only a few hundred femtoseconds \cite{Wang2016}. Nonlocal Hanle measurements, meanwhile, have revealed spin lifetimes up to a few tens of picoseconds \cite{Omar2017, Yan2016, Dankert2017} that can be tuned by a back gate \cite{Yan2016, Dankert2017}. Finally, charge transport measurements on a Hall bar demonstrated a large nonlocal signal that was related to the spin Hall effect \cite{Avsar2014}. Fits to experimental measurements have estimated the induced SOC in graphene to be 10-20 meV \cite{Avsar2014, Wang2016}, while most density functional theory (DFT) and tight-binding (TB) calculations find values closer to 1 meV \cite{Kaloni2014, Wang2015b, Yang2016, Gmitra2015, Gmitra2016, Alsharari2016}. While these studies have demonstrated that TMDCs induce strong SOC in graphene, the estimated values of $\tau_s$ vary by three orders of magnitude and nothing is yet known about the mechanisms governing spin dynamics and relaxation in these systems.

In this Letter, we employ dissipative quantum spin dynamics arguments, and quantum mechanical numerical simulations, to elucidate the nature of spin relaxation in graphene/TMDC heterostructures. We find that spin relaxation follows the D'yakonov-Perel' (DP) mechanism, with $\tau_s = 1-100$ ps for realistic momentum relaxation rates and Fermi energies. Remarkably, the spin lifetime anisotropy, defined as the ratio of lifetimes for spins pointing out of the graphene plane to those pointing in the plane, can reach unprecedented values of tens to hundreds in the presence of intervalley scattering. This behavior is mediated by spin-valley locking induced in graphene by the TMDC, which ties the in-plane spin lifetime to the intervalley scattering time. In the absence of valley mixing this ratio reduces to 1/2, typical of systems dominated by Rashba SOC \cite{Fabian2007}. A giant spin lifetime anisotropy thus represents a qualitatively new regime of spin relaxation not typically seen in 2D systems, and its measurement \cite{Raes2016, Raes2017} should be an experimental probe of systems with strong spin-valley coupling, which includes both graphene/TMDC heterostructures and TMDCs themselves. Furthermore, systems with giant spin lifetime anisotropy could serve as an exciting new platform for the manipulation of spin and the implementation of new spintronic devices.

\textit{Dissipative spin dynamics model}. To clarify the nature of spin relaxation in graphene/TMDC systems, we follow the approach in \cite{Fabian2007}, which describes spin dynamics in a randomly fluctuating magnetic field. The low-energy ($E_F < 300$ meV) Hamiltonian of graphene on a TMDC substrate is given by $H = H_0 + H_\Delta + H_I^{A/B} + H_R + H_{PIA}^{A/B}$, where \cite{Gmitra2016}
\begin{gather}
H_0 = \hbar v_F (\kappa \sigma_x k_x + \sigma_y k_y), \nonumber \\
H_\Delta = \Delta \sigma_z, \nonumber \\
H_I^{A/B} = \frac{1}{2} [\lambda_I^A (\sigma_z + \sigma_0) + \lambda_I^B (\sigma_z - \sigma_0)]\kappa s_z, \label{eq:h_cont1} \\
H_{PIA}^{A/B} = \frac{a}{2} [\lambda_{PIA}^A (\sigma_z + \sigma_0) + \lambda_{PIA}^B (\sigma_z - \sigma_0)] (k_x s_y - k_y s_x), \nonumber \\
H_R = \lambda_R (\kappa \sigma_x s_y - \sigma_y s_x).  \nonumber
\end{gather}
In Eq.\ (\ref{eq:h_cont1}), $v_F$ is the Fermi velocity, $\kappa = 1(-1)$ for the K (K$'$) valley, $\sigma_i$ ($s_i$) are the sublattice (spin) Pauli matrices, $k_i$ are the wave vector components relative to K or K$'$, and $a = 0.246$ nm is the graphene lattice constant. $H_0$ represents the graphene Dirac cone, and $H_\Delta$ is a staggered sublattice potential induced by the TMDC. $H_I^{A/B}$ and $H_{PIA}^{A/B}$ are the intrinsic and the pseudospin inversion asymmetry (PIA) SOC, respectively, the latter of which is permitted by broken z/-z symmetry in graphene \cite{Kochan2017}. Due to the broken sublattice symmetry, these terms can have different strengths and signs on the A and B sublattices ($\lambda_I^{A/B}$ and $\lambda_{PIA}^{A/B}$). Finally, $H_R$ is the Rashba SOC induced by a perpendicular electric field \cite{Kane2005, Gmitra2009}.

\begin{figure}[t]
\includegraphics[width=\columnwidth]{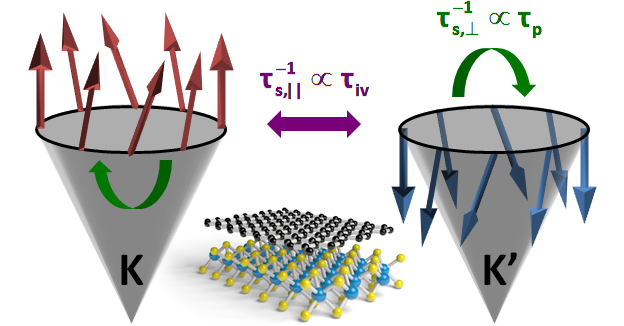}
\caption{Schematic of spin relaxation in graphene/TMDC heterostructures. The tall arrows depict the effective spin-orbit field within the Dirac cones at K and K$'$ valleys. Intervalley scattering dominates the in-plane spin dynamics, while overall momentum scattering controls the out-of-plane behavior.}
\label{fig:schematic}
\end{figure}

While Eq.\ (\ref{eq:h_cont1}) is useful for TB calculations, analytically it is more transparent to combine the sublattice-dependent terms, giving $H = H_0 + H_\Delta + H_I + H_{VZ} + H_R + H_{PIA} + H_{\Delta_{PIA}}$ with
\begin{gather}
H_I = \lambda_I \kappa \sigma_z s_z, \nonumber \\
H_{VZ} = \lambda_{VZ} \kappa s_z, \label{eq:h_cont2} \\
H_{PIA} = a \lambda_{PIA} \sigma_z (k_x s_y - k_y s_x), \nonumber \\
H_{\Delta_{PIA}} = a \Delta_{PIA} (k_x s_y - k_y s_x), \nonumber
\end{gather}
where $\lambda_I = (\lambda_I^A + \lambda_I^B)/2$, $\lambda_{VZ} = (\lambda_I^A - \lambda_I^B)/2$, $\lambda_{PIA} = (\lambda_{PIA}^A + \lambda_{PIA}^B)/2$, and $\Delta_{PIA} = (\lambda_{PIA}^A - \lambda_{PIA}^B)/2$. In this form, $H_I$ is the usual intrinsic SOC in graphene, which opens a topological gap 2$\lambda_I$ at the Dirac point \cite{Kane2005}. $H_{VZ}$ is a valley Zeeman term, which locks valley to spin and polarizes the bands out of the graphene plane with opposite orientation in the K and K$'$ valleys. $H_{PIA}$ renormalizes the Fermi velocity, while $H_{\Delta_{PIA}}$ leads to a $k$-linear splitting of the bands, as in traditional 2D electron gases with Rashba SOC \cite{Rashba1984}. Except for the PIA terms, this Hamiltonian is the same as that considered in previous works \cite{Wang2015b, Yang2016, Wang2016, Alsharari2016}.

The next step is to derive the effective spin-orbit field arising from the SOC terms. This is done by rewriting Eq.\ (\ref{eq:h_cont2}) in the basis of the eigenstates of $H_0$ and projecting onto the conduction and valence bands. At Fermi energies away from the Dirac point ($E_F \gg 1$ meV), this gives
\begin{gather}
H = H_0 + \frac{1}{2} \hbar \vec{\omega}(t) \cdot \vec{s}, \nonumber \\
\hbar \omega_x = -2(ak\Delta_{PIA} \pm \lambda_R) \sin \theta, \label{eq:h_cont3} \\
\hbar \omega_y =  2(ak\Delta_{PIA} \pm \lambda_R) \cos \theta, \nonumber \\
\hbar \omega_z = 2 \kappa \lambda_{VZ}, \nonumber
\end{gather}
where $k$ is the wave vector magnitude, $\theta$ is the direction of $k$ with respect to $k_x$, and $\vec{\omega}$ is the spin precession frequency of the effective spin-orbit field. The in-plane components of $\vec{\omega}$ give a Rashba-like spin texture, where $+(-)$ is for the conduction (valence) band. Strong PIA SOC thus leads to electron-hole asymmetry, as will be seen for graphene on WS$_2$. The out-of-plane component of $\vec{\omega}$ is determined by $\lambda_{VZ}$ and changes sign between valleys. The overall texture of the effective spin-orbit field is depicted in Fig.\ \ref{fig:schematic}.

Owing to momentum scattering, each component of $\vec{\omega}$ will fluctuate in time. A simple model for the correlation of the fluctuating spin-orbit field is \cite{Fabian2007}
\begin{gather}
\overline{\omega_\alpha(t) \omega_\beta(t')} = \delta_{\alpha \beta} \overline{\omega_\alpha^2} \mathrm{e}^{-|t-t'|/\tau_{c,\alpha}}, \label{eq:correlation}
\end{gather}
where the correlation time of fluctuation $\tau_{c,\alpha}$ depends on the component of $\vec{\omega}$. The in-plane components $\omega_{x/y}$ depend only on $\theta$, and thus $\tau_{c,x} = \tau_{c,y} = \tau_p$, the momentum relaxation time. Meanwhile, the out-of-plane component $\omega_z$ depends only on the valley index, giving $\tau_{c,z} = \tau_{iv}$, the intervalley scattering time. Assuming that $\tau_{c,\alpha} \omega_\alpha \ll 1$, applying Eqs. (\ref{eq:h_cont3}) and (\ref{eq:correlation}) to the equation of motion for the density matrix \cite{Fabian2007} yields the final expressions for the spin relaxation rates
\begin{gather}
\tau_{s,x}^{-1} = \overline{\omega_z^2} \tau_{iv} + \overline{\omega_y^2} \tau_p, \nonumber \\
\tau_{s,y}^{-1} = \overline{\omega_z^2} \tau_{iv} + \overline{\omega_x^2} \tau_p, \label{eq:rates_iv} \\
\tau_{s,z}^{-1} = (\overline{\omega_x^2} + \overline{\omega_y^2}) \tau_p. \nonumber
\end{gather}
In Eq.\ (\ref{eq:rates_iv}), the out-of-plane spin relaxation follows the usual DP relation, $\tau_{s,\perp}^{-1} \equiv \tau_{s,z}^{-1} = [2(ak\Delta_{PIA} \pm \lambda_R) / \hbar]^2 \tau_p$, with the Rashba SOC augmented by the PIA term. However, the in-plane relaxation includes contributions from both the intervalley and the overall momentum scattering, and is given by $\tau_{s,\parallel}^{-1} \equiv \tau_{s,x}^{-1} = \tau_{s,y}^{-1} = (2 \lambda_{VZ} / \hbar)^2 \tau_{iv} + \tau_{s,z}^{-1}/2$. The nature of the spin relaxation, with $\tau_{s,\parallel}$ determined by $\tau_{iv}$ and $\tau_{s,\perp}$ by $\tau_p$, is shown schematically in Fig.\ \ref{fig:schematic}. Ignoring the PIA term, the spin lifetime anisotropy is
\begin{gather}
\frac{\tau_{s,\perp}} {\tau_{s,\parallel}} = \left( \frac{\lambda_{VZ}} {\lambda_R} \right)^2 \left( \frac{\tau_{iv}} {\tau_p} \right) + 1/2. \label{eq:anisotropy}
\end{gather}
Equation (\ref{eq:anisotropy}) is the main result of this work, and indicates that a giant spin lifetime anisotropy, with the in-plane spins relaxing much faster than the out-of-plane spins, should be a defining characteristic of systems with strong spin-valley locking. Using DFT values of $\lambda_{VZ} = 1.2$ meV and $\lambda_R = 0.56$ meV for graphene on WSe$_2$ \cite{Gmitra2016}, and assuming relatively strong intervalley scattering ($\tau_{iv} \sim 5 \tau_p$), we obtain a spin lifetime anisotropy of $\sim$20. This represents a qualitatively different regime of spin relaxation than the usual case of 2D Rashba systems, where without valley Zeeman SOC the anisotropy is 1/2, with the in-plane spins relaxing more slowly than the out-of-plane spins.

Equation (\ref{eq:rates_iv}) assumes strong intervalley scattering, $\tau_{iv} \omega_z \ll 1$, such that fast fluctuation of $\omega_z$ results in motional narrowing of the in-plane spin precession and an inverse dependence of $\tau_{s,\parallel}$ on $\tau_{iv}$. In contrast, when $\tau_{iv} \rightarrow \infty$, electrons experience a constant out-of-plane spin-orbit field and only the in-plane components fluctuate with time. In this limit, the procedure above yields
\begin{gather}
\tau_{s,x}^{-1} = \overline{\omega_y^2} \tau_p^*, \nonumber \\
\tau_{s,y}^{-1} = \overline{\omega_x^2} \tau_p^*, \label{eq:rates_noiv} \\
\tau_{s,z}^{-1} = (\overline{\omega_x^2} + \overline{\omega_y^2}) \tau_p^*, \nonumber
\end{gather}
where $\tau_p^* = \tau_p / (\omega_z^2 \tau_p^2 + 1)$. Without intervalley scattering the spin lifetime anistropy thus collapses to $1/2$, as found in Rashba systems \cite{Fabian2007}. Interestingly, in this regime an external perpendicular magnetic field $B_z$ can induce an imbalance in the spin population of each valley by enhancing (canceling) the spin-orbit field at K (K$'$). The ratio of spin lifetimes in each valley thus becomes
\begin{gather}
\frac{\tau_{s,\alpha}^K}{\tau_{s,\alpha}^{K'}} = \frac{(g\mu_B B_z + 2\lambda_{VZ})^2 \tau_p^2 + 1}{(g\mu_B B_z - 2\lambda_{VZ})^2 \tau_p^2 + 1}, \label{eq:ts_ratio} 
\end{gather}
where $g$ is the electron g-factor and $\mu_B$ is the Bohr magneton. For graphene on WSe$_2$ with $\tau_p = 100$ fs, the difference in $\tau_s$ can reach 10\% for $B_z \approx 4$ T. Although this difference is too modest to achieve a complete valley-spin imbalance, it should be considered when observing spin relaxation in these structures in a magnetic field.

\begin{figure}[t]
\includegraphics[width=\columnwidth]{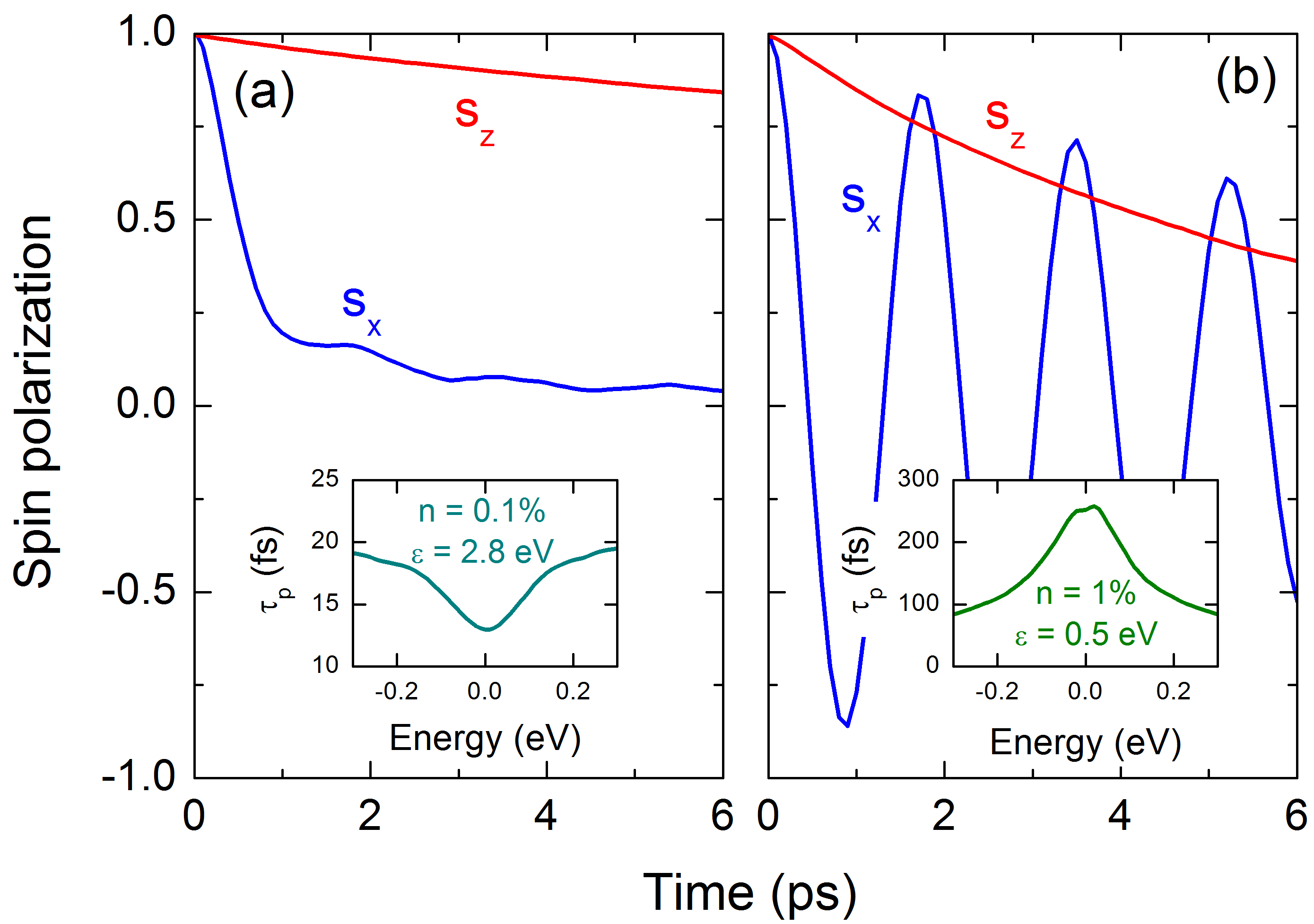}
\caption{Spin dynamics in the graphene/WSe$_2$ system for (a) strong and (b) weak intervalley scattering. The insets show the corresponding momentum relaxation times.}
\label{fig:svst}
\end{figure}

\textit{Numerical simulations}. To verify the above results, we perform numerical simulations of spin relaxation. The graphene/TMDC system is described by the TB form of Eq.\ (\ref{eq:h_cont1}), to which we add a disorder term $H_{dis} = \sum \nolimits_{i,s} V_{dis}(\vec{r}_i) \, c_{i s}^\dagger c_{i s}$, where $c_{i s}^\dagger (c_{i s})$ is the creation (annihilation) operator at site $i$ with spin $s$, and $V_{dis}(\vec{r}_i)$ is the potential at site $i$. We assume the disorder consists of Gaussian-shaped electron-hole puddles \cite{Adam2011}, such that $V_{dis}(\vec{r}_i) = \sum \nolimits_{j=1}^N \epsilon_j \mathrm{exp}(-|\vec{r}_i - \vec{r}_j|^2 / 2\xi^2)$, with the strength $\epsilon_j$ of each scatterer randomly chosen within $[-\epsilon,\epsilon]$, and with a uniform width $\xi = \sqrt{3}a$. In the dilute limit, $\tau_p$ and $\tau_{iv}$ are inversely proportional to the number of scatterers $N$, while $\epsilon$ controls their relative magnitude, with larger $\epsilon$ giving stronger intervalley scattering \cite{Zhang2009, Ortmann2011}.

To calculate charge and spin transport, we employ a real-space wavepacket propagation method that allows for efficient simulation of large-scale disordered graphene systems \cite{Roche1999,Cummings2014, VanTuan2016}. For charge transport we use the mean-square spreading of the wavepacket $\langle X^2(E,t) \rangle$ to calculate the diffusion coefficient $D(E,t) = \partial \langle X^2(E,t) \rangle / \partial t$, which in turn gives the momentum relaxation time $\tau_p(E) = \max D(E,t) / 2v_F^2$. We simultaneously calculate the expectation value of the spin of the wavepacket $\vec{s}(E,t)$, from which the spin lifetime is evaluated by fitting to $\exp(-t/\tau_{s,\alpha})$ or $\exp(-t/\tau_{s,\alpha})\cos(\omega_z t)$, as appropriate. The density of charge scatterers is characterized as a percentage of the number of carbon atoms, $n = N/N_C \times 100\%$. We consider a 500 nm $\times$ 500 nm system with 9.2 million carbon atoms, and TB parameters are taken from Table I of Ref.\ \cite{Gmitra2016}.

\begin{figure}[t]
\includegraphics[width=\columnwidth]{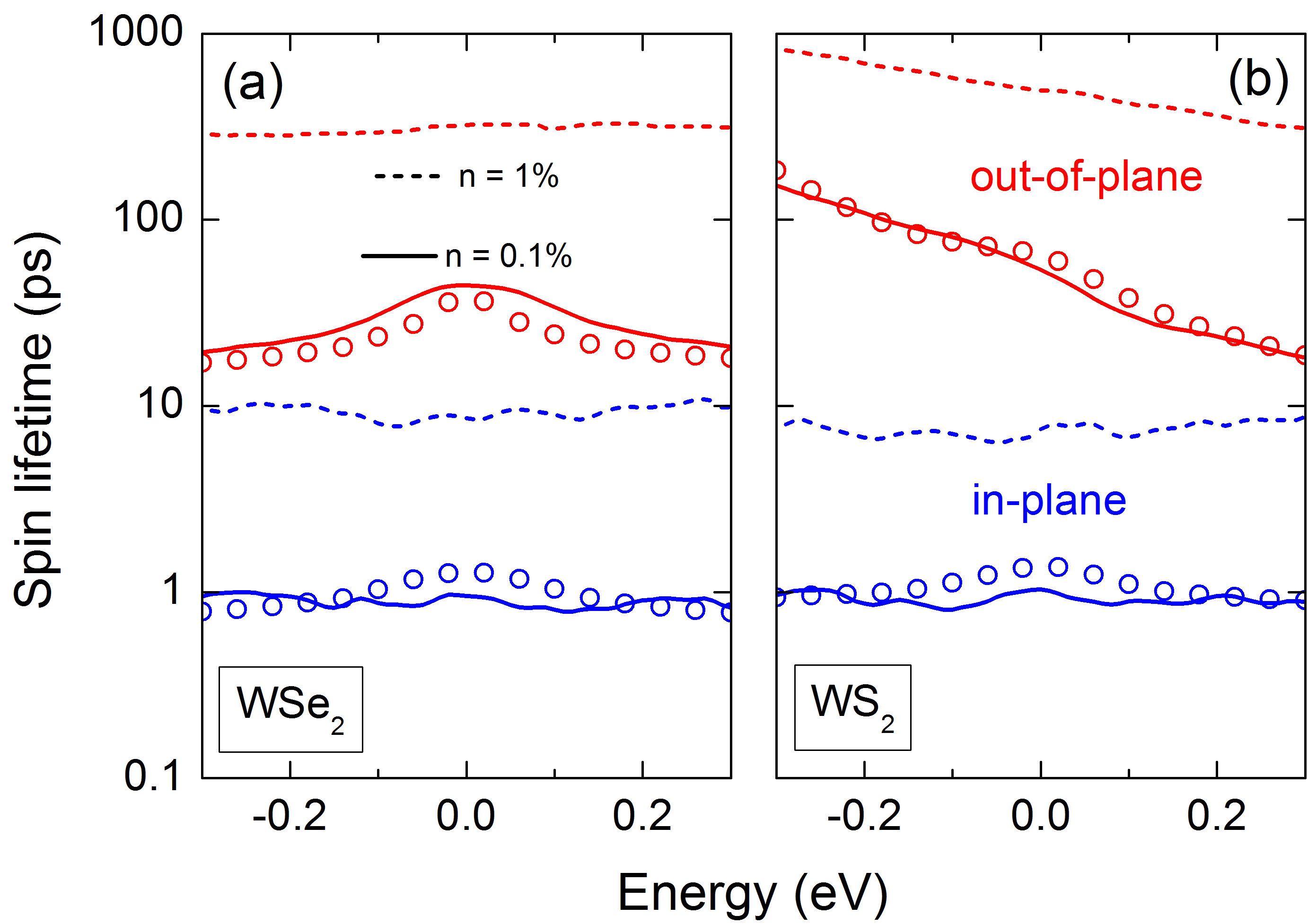}
\caption{Spin lifetime with strong intervalley scattering for graphene on (a) WSe$_2$ and (b) WS$_2$. The red (blue) lines are for out-of-plane (in-plane) spin lifetime. Solid (dashed) lines are for an impurity density of 0.1\% (1\%). The open circles are from Eq.\ (\ref{eq:rates_iv}).}
\label{fig:ts_iv}
\end{figure}

Figures \ref{fig:svst}(a) and (b) show $\vec{s}$ and $\tau_p$ for disorder profiles corresponding to strong and weak intervalley scattering, respectively. In the former we set $n = 0.1\%$ and $\epsilon = 2.8$ eV, and in the latter $n = 1\%$ and $\epsilon = 0.5$ eV. The $\tau_p$ for these two cases are shown in the insets, with values typical of those found experimentally \cite{Yang2016, Wang2015b, Wang2016}. The different energy dependence of $\tau_p$, with a minimum or maximum at the Dirac point, is indicative of the contribution of intervalley scattering \cite{Roche2012}. In Fig.\ \ref{fig:svst}(a), where intervalley scattering is strong, the in-plane component of $\vec{s}$ decays much more quickly than the out-of-plane component, and spin precession is suppressed. Meanwhile, in Fig.\ \ref{fig:svst}(b) the in-plane spin precesses about the effective spin-orbit field with frequency $\omega_z = 2\lambda_{VZ}/\hbar$, and relaxes more slowly than the out-of-plane spin. This behavior is consistent with Eqs. (\ref{eq:rates_iv})-(\ref{eq:rates_noiv}).

Figure \ref{fig:ts_iv} shows the numerical spin lifetimes in the case of strong intervalley scattering for graphene on (a) WSe$_2$ and (b) WS$_2$. The solid lines, for $n = 0.1\%$, indicate a giant anisotropy with $\tau_{s,\perp} = 20-200$ ps and $\tau_{s,\parallel} \approx 1$ ps. There is also a significant electron-hole asymmetry in $\tau_{s,\perp}$ for graphene on WS$_2$, arising from the larger PIA SOC in this system; $\lambda_R = 0.56$ meV and $\Delta_{PIA} = 75$ $\mu$eV for WSe$_2$, while $\lambda_R = 0.36$ meV and $\Delta_{PIA} = 1.4$ meV for WS$_2$ \cite{Gmitra2016}. The open circles are the values of $\tau_s$ estimated from Eq.\ (\ref{eq:rates_iv}), showing good agreement between the numerical simulations and the spin dynamics model. To fit $\tau_{s,\parallel}$ we assumed $\tau_{iv} = 5\tau_p$; although our calculations do not permit an exact determination of $\tau_{iv}$, this ratio is consistent with prior numerical results \cite{Zhang2009}. As shown by the dashed lines, increasing the disorder density to $n = 1\%$ scales $\tau_s$ by a factor of 10, confirming the inverse relationship between $\tau_s$ and $\tau_{p,iv}$.

\begin{figure}[t]
\includegraphics[width=\columnwidth]{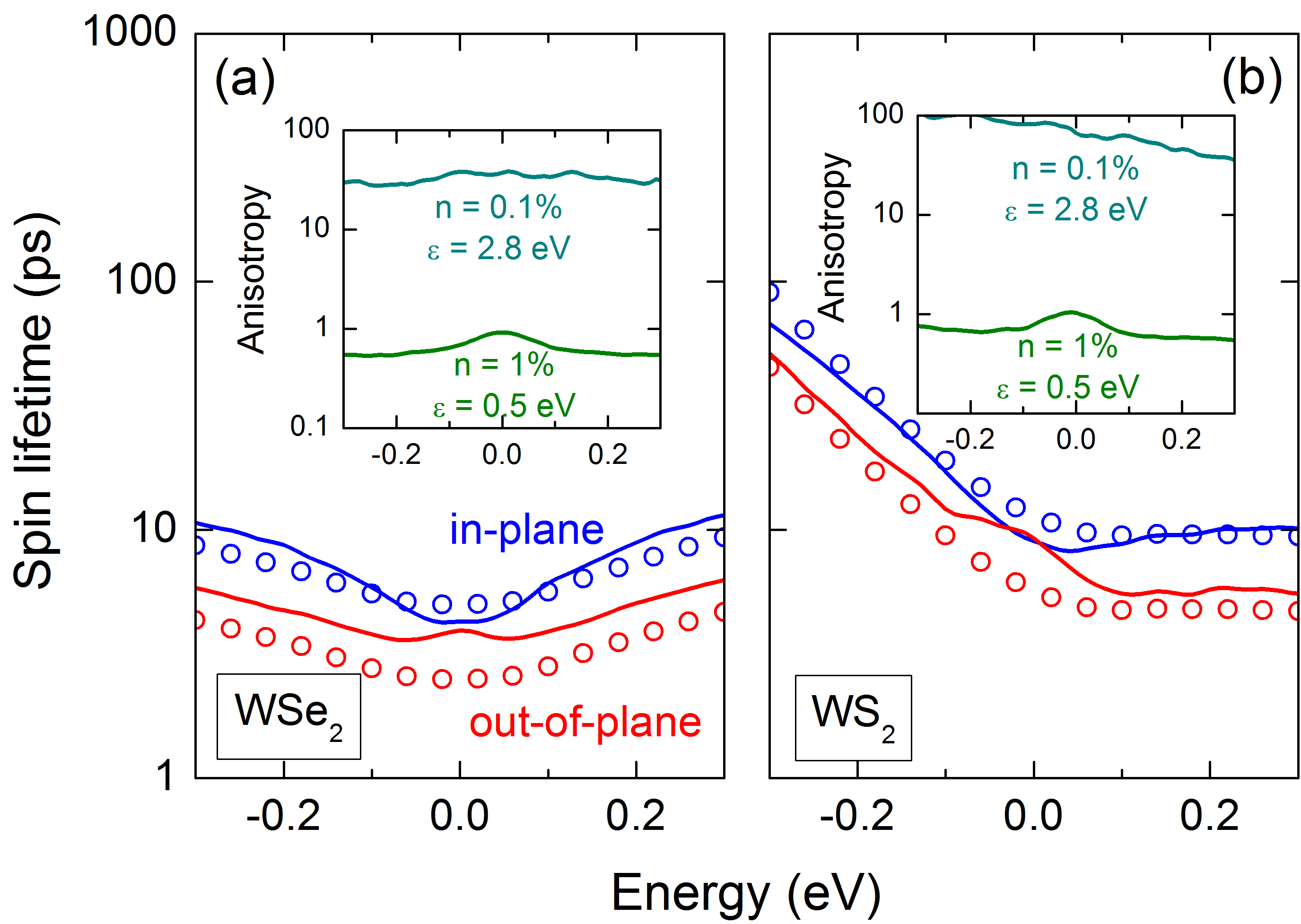}
\caption{Spin lifetime without intervalley scattering for graphene on (a) WSe$_2$ and (b) WS$_2$. The red (blue) lines are for out-of-plane (in-plane) spin lifetime. The open circles are from Eq.\ (\ref{eq:rates_noiv}), and the insets show the anisotropy for strong and weak intervalley scattering.}
\label{fig:ts_noiv}
\end{figure}

The numerical spin lifetimes in the absence of intervalley scattering are shown in Fig.\ \ref{fig:ts_noiv}, where $\tau_{s,\parallel}$ is now larger than $\tau_{s,\perp}$. The agreement with the predictions of Eq.\ (\ref{eq:rates_noiv}), shown as the open circles, is very convincing. However, we note that the agreement worsens at low energies, as the effective spin-orbit field in Eq.\ (\ref{eq:h_cont3}) is only valid for energies away from the Dirac point. The insets of Fig.\ \ref{fig:ts_noiv} show the numerical values of the spin lifetime anistropy. As predicted by the semiclassical theory, the anisotropy is giant in the case of strong intervalley scattering, and collapses toward 1/2 otherwise.

\textit{Summary and conclusions}. Using realistic quantum spin dynamics modeling and numerical simulations, we have presented a unified picture of the spin relaxation in graphene on TMDCs. We predict a giant spin relaxation anisotropy, which emerges in graphene due to proximity effects but should exist in any system with strong spin-valley locking, including TMDCs themselves. In the absence of spin-valley locking or intervalley scattering the anisotropy falls to 1/2, as expected for Rashba systems. This large variation indicates a qualitatively new regime of spin relaxation in graphene and other 2D materials.

It should be noted that the theory presented here is applicable when spin relaxation is dominated by SOC, but other spin relaxation mechanisms can take over when the SOC is small. This appears to be the case for graphene on SiO$_2$, where measurements yielded no anisotropy, i.e., $\tau_{s,\perp} = \tau_{s,\parallel}$ \cite{Raes2016, Raes2017}. In these systems the SOC is small and spin relaxation is likely dominated by paramagnetic impurities \cite{Kochan2014, Soriano2015}. Meanwhile, very recent measurements have confirmed our prediction of giant spin lifetime anisotropy in graphene/TMDC heterostructures, with an anisotropy of $\sim$11 (40) for graphene on MoSe$_2$ (WSe$_2$) at a temperature of 75 K \cite{Ghiasi2017}. Another recent measurement found an anisotropy of $\sim$10 in graphene on WS$_2$ at room temperature \cite{Benitez2017}, suggesting that temperature-dependent effects driven by electron-electron or electron-phonon scattering should have a weak impact.

These results also have important implications for the WAL analysis of magnetotransport in graphene/TMDC heterostructures. Previous analyses have concluded that the spin relaxation is dominated by Rashba SOC \cite{Yang2016, Yang2017}, which is seemingly at odds with the presence of giant spin lifetime anisotropy. By reanalyzing the magnetoconductance measurements of Ref.\ \cite{Yang2017}, and introducing valley Zeeman SOC into the analysis, the experimental results can be shown to be consistent with our theory \footnote{See Supplemental Material for an analysis of the magnetoconductance measurements of Ref.\ \cite{Yang2017}. This Supplemental Material also includes Refs. \cite{McCann2012,Ochoa2012}}.

On the more applied side, the giant spin lifetime anisotropy in graphene/TMDC heterostructures might be utilized for practical purposes in spin logic devices \cite{Dery2012, Wen2016} or in relation with opto-valleytronic spin injection in graphene/TMDC spin valves \cite{Luo2017, Avsar2017}. One possible application would be the design of a linear spin polarizer, where the in-plane components of an incoming spin current would be filtered out, leaving only the net out-of-plane polarization.

\begin{acknowledgments}
ICN2 is supported by the Severo Ochoa program from Spanish MINECO (Grant No.\ SEV-2013-0295) and funded by the CERCA Programme / Generalitat de Catalunya. A. W. C., J. H. G., and S. R. acknowledge the Spanish Ministry of Economy and Competitiveness and the European Regional Development Fund (Project No.\ FIS2015-67767-P MINECO/FEDER), the Secretar\'{i}a de Universidades e Investigaci\'{o}n del Departamento de Econom\'{i}a y Conocimiento de la Generalidad de Catalunya (2014 SGR 58), PRACE and the Barcelona Supercomputing Center (Project No. 2015133194). J. F. acknowledges support from DFG SFB 1277 projects A09 and B07. All authors acknowledge the EU Seventh Framework Programme under Grant Agreement No. 696656 Graphene Flagship.
\end{acknowledgments}

\bibliography{grtmdc_spin_aniso_bib}

\end{document}


\title{Supplementary Material:\\Giant Spin Lifetime Anisotropy in Graphene Induced by Proximity Effects}

\author{Aron W. Cummings}
\email{aron.cummings@icn2.cat}
\affiliation
{
	Catalan Institute of Nanoscience and Nanotechnology (ICN2), CSIC and BIST,
	Campus UAB, Bellaterra, 08193 Barcelona, Spain
}

\author{Jose H. Garcia}
\affiliation
{
	Catalan Institute of Nanoscience and Nanotechnology (ICN2), CSIC and BIST,
	Campus UAB, Bellaterra, 08193 Barcelona, Spain
}

\author{Jaroslav Fabian}
\affiliation
{
	Insitute for Theoretical Physics,
	University of Regensburg,
	93040 Regensburg, Germany
}

\author{Stephan Roche}
\email{stephan.roche@icn2.cat}
\affiliation
{
	Catalan Institute of Nanoscience and Nanotechnology (ICN2), CSIC and BIST,
	Campus UAB, Bellaterra, 08193 Barcelona, Spain
}
\affiliation
{
	ICREA, Instituci\'{o} Catalana de Recerca i Estudis Avan\c{c}ats,
	08070 Barcelona, Spain
}
\date{\today}

\maketitle

In this Supplemental Material, we present a brief complementary analysis of the weak antilocalization (WAL) measurements of Ref.\ \cite{Yang2017} (Ref.\ [30] in the main text). This analysis is intended to show that the theory of spin lifetime anisotropy presented in this Letter can be consistent with prior measurements of WAL in graphene/TMDC heterostructures. It also underscores the need for further theoretical and experimental study of the relationship between spin transport and quantum conductivity in these systems.

We analyze the magnetoconductance (MC) measured in a graphene/WSe$_2$ heterostructure, shown in Fig.\ 3(a) of Ref.\ \cite{Yang2017}. We consider the MC taken at a gate voltage of $V_G = -32$ V, which is reproduced in the symbols of Fig.\ \ref{figS1} below. The MC exhibits a large peak around zero magnetic field, indicative of WAL induced by strong Rashba spin-orbit coupling (SOC). The data can be fit with the theory of McCann and Fal'ko \cite{McCann2012},
\begin{equation} \label{eqWAL}
\begin{split}
\Delta \sigma(B) = &- \frac{1}{2\pi} \left[ F \left( \frac{\tau_B^{-1}}{\tau_{\phi}^{-1}} \right)
- F \left( \frac{\tau_{B}^{-1}}{\tau_{\phi}^{-1}+2\tau_{asy}^{-1}} \right) \right. \\
&- 2F \left. \left( \frac{\tau_{B}^{-1}}{\tau_{\phi}^{-1}+\tau_{asy}^{-1}+\tau_{sym}^{-1}} \right) \right],
\end{split}
\end{equation}
where $\Delta \sigma$ is the quantum correction to the conductivity in units of $e^2/h$, $F(z) = \ln(z) + \Psi(1/2 + 1/z)$, $\Psi$ is the digamma function, $\tau_{B}^{-1} = 4DeB/\hbar$, $D$ is the diffusion coefficient, $\tau_\phi$ is the dephasing time, and $B$ is the external magnetic field. The spin-orbit time $\tau_{asy}$ arises from Rashba SOC, while $\tau_{sym}$ is typically assigned to intrinsic, or Kane-Mele, SOC.

The dashed line in Fig.\ \ref{figS1} shows a fit to the experimental data using Eq.\ (\ref{eqWAL}). For this fit we set $D = \mu \hbar v_F \sqrt{\pi n} / 2e$, where $\mu$ is the electron mobility, $n$ is the charge density, and $v_F = 10^8$ cm/s is the graphene Fermi velocity. Using $\mu = 10^4$ cm$^2$/V-s and $n = 2 \times 10^{12}$ cm$^{-2}$ gives $D = 0.08$ m$^2$/s. Fitting the rest of the parameters then yields $\tau_\phi = 20$ ps, $\tau_{asy} = 13$ ps, and $\tau_{sym} = 0.6$ ps.

\begin{figure}[t]
\includegraphics[width=\columnwidth]{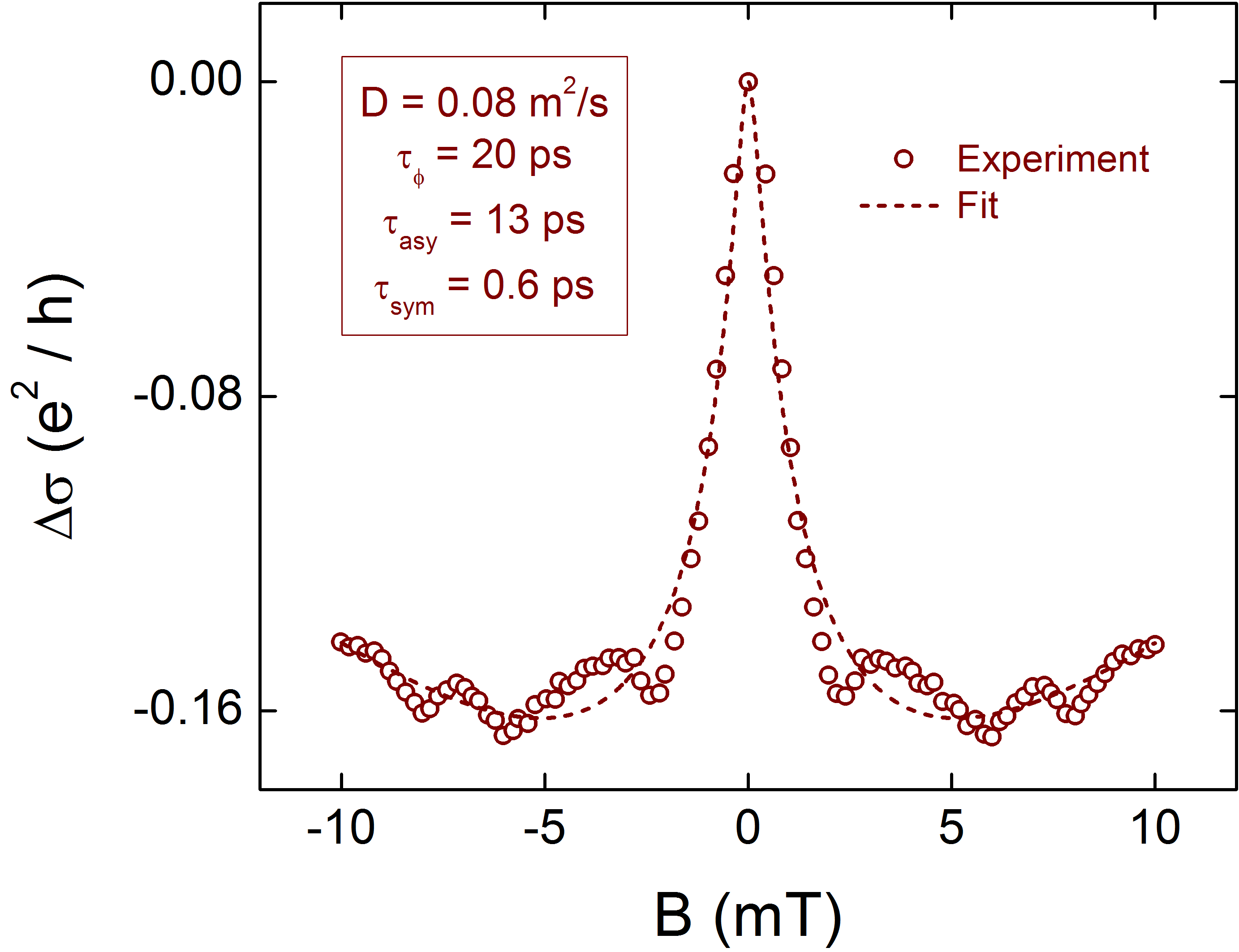}
\caption{Magnetoconductance of a graphene/WSe$_2$ heterostructure. Symbols are experimental data taken from Fig.\ 3(a) of Ref.\ \cite{Yang2017}, for $V_G = -32$ V. The dashed line is the fit using the WAL theory of Eq.\ (\ref{eqWAL}).}
\label{figS1}
\end{figure}

The first thing to consider from this fit is the magnitude of $\tau_{asy}$. Assuming this arises solely from Rashba SOC, and assuming the D'yakonov-Perel' mechanism of spin relaxation, $\tau_{asy}^{-1} = (2 \lambda_R / \hbar)^2 \tau_p$, we can estimate the Rasbha SOC strength as $\lambda_R = \hbar / \sqrt{4 \tau_{asy} \tau_p}$. For a momentum relaxation time $\tau_p \approx 100$ fs (cf. Fig.\ 4(b) of Ref.\ \cite{Yang2017}), we get $\lambda_R \approx 0.3$ meV. This rough estimate is right in the range of the values predicted by DFT \cite{Gmitra2016}.

Next we consider $\tau_{sym}$, which is usually assumed to arise from the Elliot-Yafet (EY) mechanism of spin relaxation induced by intrinsic SOC, $\tau_{sym}^{-1} = (\lambda_I / 2E_F)^2 \tau_p^{-1}$, where $E_F$ is the Fermi energy \cite{Ochoa2012}. Taking $\tau_{sym} = 0.6$ ps, $\tau_p \approx 100$ fs, and $E_F \approx 160$ meV (corresponding to $n \approx 2 \times 10^{12}$ cm$^{-2}$) gives $\lambda_I = 2E_F \sqrt{\tau_p / \tau_{sym}} \approx 130$ meV, which is an unreasonably large value. The DFT simulations of Ref.\ \cite{Gmitra2016} predict $\lambda_I$ on the order of tens of $\mu$eV, and those of Ref.\ \cite{Wang2015b} found it to be vanishingly small, but even a value of $\lambda_I = 1$ meV would give $\tau_{sym} \approx 10$ ns, four orders of magnitude larger than what is found in the above fit. This analysis thus shows that the small value of $\tau_{sym}$ does not arise from intrinsic SOC.

Instead, we posit that $\tau_{sym}$ is governed by the valley Zeeman SOC, such that $\tau_{sym}^{-1} = (2 \lambda_{VZ} / \hbar)^2 \tau_{iv}$. In Ref.\ \cite{Yang2017} it was argued that this term does not relax the spin, but we have shown that it does relax the in-plane spin. Indeed, it should be noted that the EY mechanism induced by intrinsic SOC is also an in-plane spin relaxation process, as shown in the Supplementary Information of Ref.\ \cite{Ochoa2012}. Thus, the valley Zeeman SOC can contribute to the MC through $\tau_{sym}$. If we choose $\tau_{iv} \approx 10 \tau_p \approx 1$ ps then $\lambda_{VZ} = \hbar / \sqrt{4 \tau_{sym} \tau_{iv}} \approx 0.4$ meV, which is a reasonable value in line with DFT simulations \cite{Gmitra2016}. Finally, assuming that $\tau_{asy}$ is driven by Rashba SOC and $\tau_{sym}$ by valley Zeeman SOC, the spin relaxation anisotropy would be $\tau_{asy} / \tau_{sym} = 13/0.6 = 22$. Therefore, the MC data appear to support the presence of giant spin lifetime anisotropy.

The linear scaling of $\tau_{SOC}^{-1} = \tau_{asy}^{-1} + \tau_{sym}^{-1}$ with $\tau_p$, shown in Fig.\ 4(b) of Ref.\ \cite{Yang2017}, was used as evidence that Rashba SOC was dominating the spin relaxation. However, this scaling can also occur if we consider both Rashba and valley Zeeman SOC. In this case, $\tau_{SOC}^{-1} = \tau_{asy}^{-1} + \tau_{sym}^{-1} = 4/\hbar^2 \cdot (\lambda_R^2 + \alpha \lambda_{VZ}^2) \tau_p$, where $\tau_{iv} = \alpha \tau_p$. From the fit in Fig.\ \ref{figS1}, this gives an effective Rasbha SOC of $\lambda_R^{eff} = \sqrt{\lambda_R^2 + \alpha \lambda_{VZ}^2} = 1.3$ meV, which is more or less what was extracted for the graphene/WSe$_2$ heterostructure of Ref.\ \cite{Yang2017}. Thus, in this instance, ignoring the valley Zeeman term could lead to an overstimate of the Rasbha SOC by a factor of $\sim$5.

In summary, the original analysis of Ref.\ \cite{Yang2017} concluded that spin relaxation in a graphene/WSe$_2$ heterostructure was dominated by Rashba SOC. This conclusion was supported by a large WAL peak in the magnetoconductance, and by the linear scaling of $\tau_{SOC}^{-1}$ with $\tau_p$. A Rashba-dominated spin relaxation implies a small spin lifetime anisotropy, which is seemingly in contradiction with our theory as well as recent measurements of large spin lifetime anisotropy in graphene/TMDC heterostructures \cite{Ghiasi2017, Benitez2017}. However, our revised analysis of the WAL results, shown above, indicates that large spin lifetime anisotropy actually could be present in the devices measured in Ref.\ \cite{Yang2017}. The discrepancy could lie in the fits of Eq.\ (\ref{eqWAL}) to the magnetoconductance. These fits have a degree of parametric freedom, and thus can lead to different conclusions depending on the extracted spin-orbit times. We also propose that the valley Zeeman SOC should be considered in any WAL analysis of graphene/TMDC heterostructures, as it is responsible for the in-plane spin relaxation. More theoretical and experimental work is clearly needed to fully reconcile the spin lifetimes extracted from WAL analysis to those from Hanle measurements.

\bibliography{grtmdc_spin_aniso_bib}